\documentclass[aps,pra,reprint,superscriptaddress,nofootinbib,longbibliography]{revtex4-2}

\usepackage{amsmath}
\usepackage{amssymb}
\usepackage{graphicx}
\usepackage{url}
\usepackage[colorlinks=true,linkcolor=blue,citecolor=blue,urlcolor=blue]{hyperref}
\hypersetup{
  pdftitle={The Input Problem: A Permanent Bottleneck for Quantum Machine Learning},
  pdfauthor={Muhammad Faryad},
  pdfsubject={quant-ph},
  pdfkeywords={quantum state preparation, data encoding, quantum machine learning,
    quantum amplitude estimation, dequantization}
}

\newcommand{\ket}[1]{\left|#1\right\rangle}

\begin{document}

\title{The Input Problem: A Permanent Bottleneck for Quantum Machine Learning}

\author{Muhammad Faryad}
\email{muhammad.faryad@lums.edu.pk}
\affiliation{Department of Physics, Syed Babar Ali School of Science and Engineering,\\
Lahore University of Management Sciences (LUMS), Lahore 54792, Pakistan}

\date{\today}

\begin{abstract}
Quantum algorithms are conventionally presented with their input state supplied for free. When the input is classical data, this convention conceals a cost that is frequently larger than the algorithm it precedes. We review what the three standard encodings, such as basis encoding, amplitude encoding, and Grover--Rudolph distribution loading, actually cost once transpiled to a hardware gate set, and argue that the resulting $\Theta(N)$ bound is a counting theorem rather than an engineering limitation that improved hardware will remove. Measured gate counts for a representative loading task are reported: an optimal library implementation requires $247$ CNOT gates at $n=8$ qubits and doubles with each additional qubit, while the classical preprocessing that produces the rotation angles requires reading the entire input vector. We show how this cost eliminates the quadratic advantage of quantum amplitude estimation for Monte Carlo integration, and argue that the same accounting constrains quantum machine learning more broadly: the strong input models that make quantum algorithms fast on classical data also enable classical dequantization, and quantum kernel methods carry a $\Theta(M^2)$ state-preparation cost for the Gram matrix that does not amortize. We explain that the efficiently preparable states, device-generated distributions, variationally learned loading, and amortized preparation are required to get advantage from quantum machine learning and close with a checklist for evaluating input-dependent advantage claims. Executable notebooks reproducing every construction and measurement discussed here are available.
\end{abstract}


\maketitle

\section{Introduction}
\label{sec:intro}
\footnote{Five code files implementing this paper are available at \url{https://maven.com/faryad/o/96456e}.}
Nearly any paper on quantum machine learning contains a line of the form: \emph{let
$\ket{\psi_{\mathbf{x}}}$ encode the data vector $\mathbf{x}$}. The algorithm then proceeds, and its
complexity is quoted in terms of what happens after that line.

The convention is harmless in the settings where it originated. In Shor's algorithm the input is a
single integer. In Grover's, it is a black-box oracle whose internal cost is explicitly excluded from
the analysis by construction. In both cases nothing is being concealed.

It becomes something else entirely when the input is a length-$N$ vector of real numbers drawn from
an actual dataset --- a feature vector, a covariance matrix, a probability distribution calibrated to
market data. Then the cost of producing $\ket{\psi_{\mathbf{x}}}$ is a cost of the algorithm. Because
that cost resides in a preparation stage that no complexity table lists and no circuit diagram
displays, it is routinely omitted from precisely the comparisons it should dominate.

This paper is concerned with what that cost is. In brief: for generic classical data it is
$\Theta(N)$, provably so, and it is therefore not a hardware problem that better qubits will
eventually solve. Any quantum algorithm consuming such data in $O(\operatorname{polylog} N)$ time is
dominated by its own input stage by a factor of $N/\operatorname{polylog} N$.

That is a strong claim, and Secs.~\ref{sec:encodings}--\ref{sec:qmc} are spent establishing it.
Section~\ref{sec:qml} argues that the accounting bears on quantum machine learning with particular
force, and Sec.~\ref{sec:survives} is spent bounding the claim, which matters as much as making it.
The input problem does not render quantum machine learning futile. It renders a particular
\emph{style} of advantage claim untenable, and distinguishing the two is the difference between
building on sand and building on rock.

\section{Three encodings and their costs}
\label{sec:encodings}

There are three standard encodings. They differ exponentially in qubit count and not at all in
asymptotic gate count, which is the first point worth internalizing.

\subsection{Basis encoding}

Each datum becomes the bit pattern of a computational basis state. A real number is written in fixed
point --- one sign bit, $t_L$ bits of integer part, $t_R$ bits of fraction --- and encoded by applying
$X$ wherever the bit is $1$. A set of $M$ patterns can be placed in equal superposition,
\begin{equation}
  \ket{\psi} = \frac{1}{\sqrt{M}}\sum_{i=0}^{M-1} \ket{x^{(i)}},
\end{equation}
by the routine of Ventura and Martinez~\cite{ventura1998}. Two ancilla qubits track which branch is
still being processed; at step $i$, with $M-i$ patterns remaining, a controlled rotation by
\begin{equation}
  \theta_i = 2\arcsin\!\left(1/\sqrt{M-i}\right)
\end{equation}
transfers amplitude
\begin{equation}
  \sqrt{\frac{M-i}{M}}\cdot\frac{1}{\sqrt{M-i}} = \frac{1}{\sqrt{M}},
\end{equation}
independently of $i$. The angle varies precisely so that the amplitude does not, and this telescoping
is the entire content of the construction.

The cost is $O(t)$ qubits with $t = 1 + t_L + t_R$, and $O(Mt)$ depth. The depth is not an artifact
of the method. Arbitrary $M$ patterns of $t$ bits carry $\Theta(Mt)$ bits of information; a circuit
of depth $d$ over a fixed gate set on $t + O(1)$ qubits is specified by $O(dt)$ bits; and the prepared
state determines the patterns. Hence $d = \Omega(M)$ for \emph{any} construction. Loading $M$ data
points costs at least $M$ steps, and an algorithm running in $o(M)$ has been charged $\Omega(M)$
before it begins.

\subsection{Amplitude encoding}

The alternative packs $2^n$ complex numbers into the amplitudes of $n$ qubits,
\begin{equation}
  \ket{\psi_{\mathbf{x}}} = \frac{1}{\lVert\mathbf{x}\rVert}\sum_{k=0}^{2^n-1} x_k \ket{k},
\end{equation}
a genuine exponential compression in qubit count and the reason amplitude encoding appears in nearly
every quantum machine learning proposal. A $1024$-dimensional feature vector occupies ten qubits.

The construction is a binary tree of uniformly controlled rotations~\cite{mottonen2004}. Writing
$a_k = |x_k|$ and adopting the convention $k = \sum_j q_j 2^j$, so that qubit $j$ carries bit $j$ of
the index, the rotation applied to qubit $i$ conditioned on the low $i$ bits reading $j$ is
\begin{equation}
  \theta_{i,j} = 2\arcsin\sqrt{\mathcal{A}_{i,j}/\mathcal{N}_{i,j}},
  \label{eq:ampangle}
\end{equation}
with
\begin{equation}
  \mathcal{N}_{i,j} = \!\!\!\sum_{k\equiv j \bmod 2^i}\!\!\! a_k^2,
  \qquad
  \mathcal{A}_{i,j} = \!\!\!\!\sum_{k\equiv j+2^i \bmod 2^{i+1}}\!\!\!\! a_k^2 .
\end{equation}
A matching cascade of controlled phase gates carries $\arg x_k$. Summing over levels,
$1 + 2 + \cdots + 2^{n-1} = 2^n - 1$ angles are required for the magnitudes, and as many again for
the phases.

That number deserves emphasis. Those angles are computed classically, on an ordinary computer, before
a single quantum gate exists. Producing them requires reading the entire data vector. Whatever the
quantum circuit does subsequently, the classical preprocessing has already touched every element of
the input.

\subsection{Distribution loading}

The third case is a probability density discretized onto $2^n$ bins,
$\ket{\psi_p} = \sum_k \sqrt{p_k}\ket{k}$ with $p_k = \int_{x_k}^{x_{k+1}} p(x)\,dx$. Grover and
Rudolph~\cite{grover2002} construct it by bisection: at level $i$, region $j$ spanning $[x_L, x_R)$
with midpoint $x_M$, define
\begin{equation}
  f(i,j) = \frac{\int_{x_L}^{x_M} p(x)\,dx}{\int_{x_L}^{x_R} p(x)\,dx},
  \label{eq:gr}
\end{equation}
and rotate by $\theta = \arccos\sqrt{f(i,j)}$. The conditional square roots telescope down the tree
to $\sqrt{p_k}$ at the leaves. This is the construction most often invoked in support of the position
that state preparation need not be expensive, and Sec.~\ref{sec:structure} examines the hypothesis on
which that position rests.

\section{The cost is a theorem, not an engineering problem}
\label{sec:bound}

\emph{Lower bound.} Preparing an arbitrary $\ket{\psi} \in \mathbb{C}^{2^n}$ from $\ket{0}^{\otimes n}$
requires $\Omega(2^n)$ two-qubit gates.

The argument is a dimension count. The target manifold of normalized states has $2^{n+1}-2$ real
parameters. A circuit of $g$ gates drawn from a fixed two-qubit gate set with continuous parameters
spans a manifold of dimension $O(g)$. Covering the target therefore requires $g = \Omega(2^n)$. The
bound is met up to constants by known constructions~\cite{mottonen2004,plesch2011,shende2006}, and it
is the construction that Qiskit's \texttt{StatePreparation} implements.

This changes the character of the problem. A $\Theta(N)$ gate count is not a limitation of current
compilers, qubits, or decompositions. Improved hardware will not remove it. It is a statement about
how much information a circuit can inject per gate, and it will be equally true of a fault-tolerant
machine decades hence.

Table~\ref{tab:counts} makes the point concretely. We prepare the discretized density
$p(x) = \tfrac{\pi}{2}\sin(\pi x)$ on $[0,1]$ by two routes --- a direct ancilla-based decomposition
of the Grover--Rudolph recursion, and Qiskit's \texttt{StatePreparation} --- and transpile both to the
hardware basis $\{\texttt{cx}, \texttt{rz}, \texttt{sx}, \texttt{x}\}$ at optimization level~1. Both
circuits were verified against the exact bin integrals to a maximum deviation below $10^{-16}$, so the
comparison is between two correct preparations of the same state.

\begin{table}[t]
\caption{\label{tab:counts}%
CNOT counts for preparing $\sum_k\sqrt{p_k}\ket{k}$ with
$p(x)=\tfrac{\pi}{2}\sin(\pi x)$, transpiled to
$\{\texttt{cx},\texttt{rz},\texttt{sx},\texttt{x}\}$ at optimization level~1. ``Angles'' counts the
rotation parameters evaluated classically before the circuit is constructed; this cost appears in no
circuit diagram and is untouched by any compiler pass.}
\begin{ruledtabular}
\begin{tabular}{rrrrrr}
$n$ & bins & angles & direct build & \texttt{StatePreparation} & ratio \\
\colrule
3 & 8   & 7   & 64      & 4   & 16 \\
5 & 32  & 31  & 1180    & 26  & 45 \\
7 & 128 & 127 & 8300    & 120 & 69 \\
8 & 256 & 255 & 20180   & 247 & 82 \\
\end{tabular}
\end{ruledtabular}
\end{table}

Two conclusions follow. The first is practical: a direct decomposition costs roughly eighty times the
library implementation, so the library should be used. The second is structural and more important.
Consider the \texttt{StatePreparation} column alone: it doubles with each added qubit. That
implementation is asymptotically optimal and it remains exponential. The factor of eighty is a
constant; the exponential belongs to both. The ``angles'' column, meanwhile, belongs to neither
circuit and will never be reduced by a compiler.

\section{Structure as the escape, and its actual width}
\label{sec:structure}

The lower bound of Sec.~\ref{sec:bound} concerns \emph{arbitrary} states. The natural question is what
structure buys, and the answer is: a great deal, under a hypothesis stronger than it first appears.

Return to Eq.~\eqref{eq:gr}. The ratio depends only on the cumulative distribution function evaluated
at three points. If the CDF is \emph{efficiently integrable} --- computable by a classical circuit of
size $\operatorname{poly}(n)$ --- then a reversible version of that circuit realizes an oracle
\begin{equation}
  U_P : \ket{j}\ket{0} \longmapsto \ket{j}\ket{f(i,j)},
\end{equation}
and a single uniformly controlled rotation per level converts the loaded value into an angle. The cost
drops to $O(n \cdot \operatorname{poly}(n))$. This is the actual content of the Grover--Rudolph result,
and it is a real one.

Consider, however, what most implementations do instead. They evaluate $f(i,j)$ on a classical
computer, obtain $2^n - 1$ numbers, and hard-code each into its own gate. This reproduces the paper's
recursion faithfully, produces exactly the correct state, and is exponential. Both versions circulate
under the same name, frequently citing the same source, and the distinction is seldom drawn.

The efficiency resides entirely in the reversible arithmetic that is rarely built. An implementation
computing its angles classically has not implemented an efficient state preparation routine, however
faithfully it follows the recursion or however exactly it reproduces the target distribution.

A further constraint is easily missed. ``Efficiently integrable'' excludes most densities one actually
wants. A lognormal --- the distribution underlying Black--Scholes option pricing, and the single most
cited application of this machinery --- has a CDF with no elementary form.

\section{Subtracting the cost from the speedup}
\label{sec:qmc}

The clearest demonstration of what the input problem does to an advantage claim comes from quantum
Monte Carlo integration, because there the claimed advantage is stated precisely enough to subtract
from.

Quantum amplitude estimation~\cite{brassard2002} estimates $\mathbb{E}[f(X)]$ to additive error
$\epsilon$ using $O(1/\epsilon)$ queries to a state-preparation oracle $\mathcal{A}$, against
$O(1/\epsilon^2)$ samples for classical Monte Carlo. This quadratic gap underlies essentially every
proposal in quantum derivative pricing and quantum risk analysis.

Inserting the cost of $\mathcal{A}$ changes the conclusion. Resolving an expectation to relative
precision $\epsilon$ requires $N \sim 1/\epsilon$ bins; in the companion notebooks, a $32$-bin
lognormal prices a European call to within $0.5\%$ of the Black--Scholes value, and the residual is
entirely discretization rather than circuit error. By the bound of Sec.~\ref{sec:bound},
$\mathcal{A}$ costs $\Theta(N)$ gates. Therefore
\begin{equation}
  \text{total cost} =
  \underbrace{O(1/\epsilon)}_{\text{QAE queries}} \times
  \underbrace{\Theta(1/\epsilon)}_{\text{cost per } \mathcal{A}}
  = O(1/\epsilon^{2}),
  \label{eq:nospeedup}
\end{equation}
which is the classical scaling. Extrapolating the measured counts of Table~\ref{tab:counts} makes this
concrete: at $\epsilon = 10^{-5}$ the pipeline requires approximately $9.6\times10^{9}$ two-qubit
gates, against $10^{10}$ classical samples --- the same number, one on hardware that does not exist
and one on a laptop.

The quadratic advantage was counted in oracle queries under the assumption that the oracle is free.
It is not free, and its cost is of exactly the order required to close the gap. This is the argument
of Herbert~\cite{herbert2021}, who also proposes a Fourier-series route that evades the obstruction.

\section{Why this constrains quantum machine learning}
\label{sec:qml}

Quantum Monte Carlo is the cleanest illustration, but quantum machine learning is where the input
problem is most consequential, for a reason worth stating explicitly.

The speedups that motivated the field were exponential and \emph{polylogarithmic in the data
dimension}. Linear-systems-based proposals promised to manipulate an $N$-dimensional vector in
$O(\operatorname{polylog} N)$ time. That promise is meaningful only if the vector can be introduced in
comparable time, and by Sec.~\ref{sec:bound}, for generic data, it cannot. An input stage costing
$\Theta(N)$ against an algorithm costing $\operatorname{polylog}(N)$ is not a constant-factor
inconvenience; it is the entire runtime.

The standard escape is to postulate a QRAM: an oracle
$\ket{i}\ket{0}\to\ket{i}\ket{x_i}$ delivering data elements in superposition at unit cost. Two
observations follow, both anticipated in Aaronson's survey of the fine print~\cite{aaronson2015}.

First, no such device exists, and constructing one requires a hardware architecture qualitatively
different from what is presently being built.

Second, and more consequentially, the assumption is not free even in principle. A strong input model
granting efficient superposition access to a structured dataset also grants efficient \emph{classical}
sampling access to the same structure. Tang's dequantization results~\cite{tang2019} exploit precisely
this: several headline exponential quantum machine learning speedups collapse to polynomial once the
classical algorithm is granted input access comparable to what the quantum algorithm was assuming
throughout. The input assumption that made the quantum algorithm fast made a classical algorithm fast
also.

The same pressure appears in kernel methods, where it is less frequently remarked. A quantum kernel
estimates $|\langle\phi(x_i)|\phi(x_j)\rangle|^2$ for pairs of data points, requiring the preparation
of feature states for both. On $M$ training points this is $\Theta(M^2)$ state preparations to fill
the Gram matrix, each with its own circuit and its own shot budget. The feature map may be of
polynomial depth and entirely benign; the loading is nevertheless per-data-point, does not amortize
across the dataset, and the quadratic count in $M$ underlies every quantum kernel result whether or
not it is reported.

The pattern is consistent across the field. Wherever an advantage is claimed against classical data,
the assumption made about introducing that data is load-bearing --- and it is usually contained in a
sentence rather than a table.

\section{What survives}
\label{sec:survives}

Overstating the foregoing would be as careless as ignoring it, so the boundaries should be drawn
precisely. The input problem does not establish that quantum machine learning is futile. It
establishes that one specific argument pattern --- classical dataset in, polylogarithmic quantum
algorithm, exponential advantage out --- does not close. The following remain intact.

\emph{Genuinely efficiently preparable states.} Where a closed-form CDF admits a
$\operatorname{poly}(n)$ reversible circuit, the Grover--Rudolph construction performs as advertised.
Such cases are real. They are rarer than the citation pattern suggests, and constructing the
arithmetic is substantial work rather than a footnote.

\emph{Device-generated data.} If the distribution is the output of a Hamiltonian simulation, a
chemistry calculation, or a variationally trained circuit, it was never a classical vector and there
is nothing to load. This is the cleanest escape and, not coincidentally, where the strongest
near-term results are found.

\emph{Learned loading.} A shallow circuit trained to approximate a distribution --- a quantum
generative adversarial network, for instance~\cite{zoufal2019} --- avoids the $\Theta(N)$ angle count
entirely, since a trained ansatz is not a hard-coded amplitude list. Whether the approximation
suffices for a given downstream task is an open and worthwhile question.

\emph{Amortization.} If a single loaded state is consumed by many independent measurements, the
preparation cost divides. Whether it divides sufficiently is an arithmetic question with a definite
answer for any particular pipeline.

\emph{Polynomial advantages.} Nothing here forbids them. What is forbidden is an exponential
advantage on generic classical input.

\section{Discussion}
\label{sec:discussion}

The methodological point generalizes well beyond state preparation.

Every construction discussed here is \emph{correct}. Each was verified against a closed-form answer to
fourteen or more significant figures. Correctness was never in question, and correctness indicated
nothing whatever about usefulness. What separated the two was a control experiment: transpile the
alternative, count the same quantities, and place both in one table. That experiment required an
afternoon and it inverted the conclusion. It is also, in our experience, the step most frequently
omitted --- because a working circuit resembles a result, whereas running the classical baseline
resembles a formality that can wait.

We therefore propose six questions to be applied before any input-dependent advantage claim, whether
one's own or another's:

\begin{enumerate}
\item How many numbers are computed classically before the circuit exists? If the answer is
  $\Theta(N)$, the entire input has already been read.
\item Were the gate counts taken after transpilation to a hardware basis? Abstract counts conceal
  multi-controlled gates completely.
\item Was the method benchmarked against the library implementation, or against nothing?
\item Was it benchmarked against the classical algorithm, or against a weaker quantum one?
\item Are state error, discretization error, and shot noise reported separately? They routinely differ
  by ten orders of magnitude, and typically only the smallest is quoted.
\item If the speedup is stated in query complexity, what does the query cost?
\end{enumerate}

Applied to the literature, most papers address two of the six.

The input problem is permanent. It is a counting argument rather than a hardware limitation, and it
will constrain quantum machine learning for as long as quantum machine learning is applied to
classical data. This is not grounds for pessimism. It is grounds for precision about which claims a
given architecture can support. The field will be better served by a smaller number of defensible
advantages than by a larger number that dissolve on contact with a gate count.

\section*{Code availability}

Five executable Jupyter notebooks reproduce every construction and measurement reported here,
verified against Qiskit 2.5.1 and qiskit-aer 0.17.2. Each numerical result is checked against a
closed-form answer with an assertion that fails if it drifts. The notebooks cover basis encoding and
the Ventura--Martinez superposition routine; amplitude encoding, including the phase cascade and a
phase-reference error that is exact on any test vector with a real first component and wrong by
$O(1)$ otherwise; Grover--Rudolph loading verified bin by bin against exact integrals; Gaussian and
lognormal densities with a European call priced against Black--Scholes; and the benchmark of
Table~\ref{tab:counts} together with the arithmetic of Eq.~\eqref{eq:nospeedup}.

They are available at \url{https://maven.com/faryad/o/96456e}.

\begin{acknowledgments}
The author thanks the participants of the Applied Quantum Machine Learning (\url{https://maven.com/faryad})cohort, whose questions
about where the data actually enters prompted this analysis.
\end{acknowledgments}


\begin{thebibliography}{99}

\bibitem{ventura1998}
D.~Ventura and T.~Martinez,
\emph{Initializing the amplitude distribution of a quantum state},
arXiv:quant-ph/9807053 (1998).

\bibitem{mottonen2004}
M.~M\"ott\"onen, J.~J.~Vartiainen, V.~Bergholm, and M.~M.~Salomaa,
\emph{Transformation of quantum states using uniformly controlled rotations},
arXiv:quant-ph/0407010 (2004).

\bibitem{grover2002}
L.~Grover and T.~Rudolph,
\emph{Creating superpositions that correspond to efficiently integrable probability distributions},
arXiv:quant-ph/0208112 (2002).

\bibitem{brassard2002}
G.~Brassard, P.~H{\o}yer, M.~Mosca, and A.~Tapp,
\emph{Quantum amplitude amplification and estimation},
Contemp. Math. \textbf{305}, 53 (2002); arXiv:quant-ph/0005055.

\bibitem{herbert2021}
S.~Herbert,
\emph{No quantum speedup with Grover--Rudolph state preparation for quantum Monte Carlo integration},
Phys. Rev. E \textbf{103}, 063302 (2021).

\bibitem{aaronson2015}
S.~Aaronson,
\emph{Read the fine print},
Nat. Phys. \textbf{11}, 291 (2015).

\bibitem{tang2019}
E.~Tang,
\emph{A quantum-inspired classical algorithm for recommendation systems},
in \emph{Proc. 51st Annual ACM SIGACT Symposium on Theory of Computing (STOC 2019)}, p.~217;
arXiv:1807.04271.

\bibitem{plesch2011}
M.~Plesch and \v{C}.~Brukner,
\emph{Quantum-state preparation with universal gate decompositions},
Phys. Rev. A \textbf{83}, 032302 (2011).

\bibitem{shende2006}
V.~V.~Shende, S.~S.~Bullock, and I.~L.~Markov,
\emph{Synthesis of quantum-logic circuits},
IEEE Trans. Comput.-Aided Des. Integr. Circuits Syst. \textbf{25}, 1000 (2006).

\bibitem{zoufal2019}
C.~Zoufal, A.~Lucchi, and S.~Woerner,
\emph{Quantum generative adversarial networks for learning and loading random distributions},
npj Quantum Inf. \textbf{5}, 103 (2019).

\end{thebibliography}
\end{document}